\documentclass[twocolumn,showpacs,showkeys,reprint,amsmath,amssymb,prl]{revtex4-1}

\usepackage{graphicx}
\usepackage{dcolumn}
\usepackage{bm}
\usepackage{tabularx}
\usepackage{amsmath}

\begin{document}

\title{Ultralow-temperature thermal conductivity of Pr$_2$Ir$_2$O$_7$: a metallic spin-liquid candidate with quantum criticality}

\author{J. M. Ni,$^1$ Y. Y. Huang,$^1$ E. J. Cheng,$^1$ Y. J. Yu,$^1$ B. L. Pan,$^1$ Q. Li,$^1$ L. M. Xu,$^2$ Z. M. Tian,$^{2,\dagger}$ and S. Y. Li$^{1,3,*}$}

\affiliation
{$^1$State Key Laboratory of Surface Physics, Department of Physics, and Laboratory of Advanced Materials, Fudan University, Shanghai 200433, China\\
 $^2$School of Physics, and Wuhan National High Magnetic Field Center, Huazhong University of Science and Technology, Wuhan 430074, China\\
 $^3$Collaborative Innovation Center of Advanced Microstructures, Nanjing 210093, China
}

\date{\today}

\begin{abstract}
The frustrated pyrochlore iridate Pr$_2$Ir$_2$O$_7$ was proposed as a metallic quantum spin liquid located at a zero-field quantum critical point. Here we present the ultralow-temperature thermal conductivity measurements on the Pr$_2$Ir$_2$O$_7$ single crystals to detect possible exotic excitations. In zero field, the thermal conductivity shows a dramatic suppression above a characteristic temperature $T_s \approx$ 0.12 K. With increasing field, $T_s$ increases and the thermal conductivity tends to saturate above $H$ = 5 T. The Wiedemann-Franz law is verified at high fields and inferred at zero field. It suggests the normal behavior of electrons at the quantum critical point, and the absence of mobile fermionic magnetic excitations. The strong suppression of thermal conductivity is attributed to the scattering of phonons by the spin system, likely the fluctuating spins. These results shed new light on the microscopic description on this novel compound.
\end{abstract}

\maketitle

Spin ice state on a frustrated pyrochlore lattice has attracted numerous interests in condensed matter physics, due to the emergent magnetic monopole excitations from the manifold of degenerate ground states \cite{Balents,monopole}. By introducing quantum fluctuations with $J_{eff}$ = 1/2 moments, quantum spin ice (QSI) states can be stabilized, exhibiting quantum electrodynamics with extra excitations like gapless photons \cite{QSI,prb04,prx11,Meng}. Yb$_2$Ti$_2$O$_7$, Tb$_2$Ti$_2$O$_7$, and Pr$_2$Zr$_2$O$_7$ are such promising QSI candidates \cite{QSI}. On the other hand, iridates with $5d$ electrons have also drawn much attention in recent years owing to the various novel quantum phases and transitions therein, which originate from the competition between spin-orbit coupling and electron-electron correlation \cite{correlated,iridate}. When these two aspects meet in the pyrochlore iridate Pr$_2$Ir$_2$O$_7$, complex phenomena and exotic phases emerge \cite{prl06,prl07,nature10,prl10,prl11,prl13,prb13,prl132,prl133,nm14,prx14,nc15,condensation}.

Pr$_2$Ir$_2$O$_7$ is a metal with the antiferromagnetic RKKY interaction of about 20 K in Pr $4f$ moments mediated by Ir $5d$ conduction electrons \cite{prl06}. The Kondo effect leads to a partial screening of the Pr $4f$ moments and gives a lower Weiss temperature of $\theta_W$ = 1.7 K \cite{prl06}. No long range magnetic order was observed down to 70 mK evidenced by the magnetic susceptibility measurement, indicating a possible metallic spin liquid ground state, or even a U(1) QSI state \cite{prl06,condensation}. A huge and anisotropic anomalous Hall effect (AHE) was probed under magnetic fields \cite{prl07,prl11}, which may be the results of the spin chirality effect on the Ir sites from the noncoplanar spin texture of Pr $4f$ moments. The observation of AHE in the absence of uniform magnetization at zero field further indicates a long-sought chiral spin liquid state in Pr$_2$Ir$_2$O$_7$ \cite{nature10}. More interestingly, a zero-field quantum critical point (QCP) was uncovered from the diverging behavior and scaling law in the Gr\"{u}neisen ratio measurement \cite{nm14}. It was also theoretically investigated as a QCP between antiferromagentic ordering and nodal non-Fermi liquid \cite{prx14}.

For such an exotic metallic spin-liquid candidate with quantum criticality, although various efforts have been made, two main issues remain to be solved in Pr$_2$Ir$_2$O$_7$. Firstly, how do the electrons behave at the QCP? In other words, will the electrons still be well-defined Landau quasiparticles? Secondly, little information is known for possible exotic magnetic excitations in Pr$_2$Ir$_2$O$_7$, probably due to the large neutron absorption cross-section of the iridium ions, and the very small size of its single crystals \cite{KF}.

Ultralow-temperature thermal conductivity measurement is an important technique to address above two issues. For the former one, the verification of the Wiedemann-Franz (WF) law $\kappa$/$\sigma$$T$ = $\pi^2$$k$$_B$$^2$/3$e^2$ = $L_0$ can be viewed as an evidence of the survival of Landau quasiparticles at the QCP. Anomalous reduction of the Lorenz ratio $L(T)/L_0$ with $L(T)$ = $\kappa$/$\sigma$$T$ has been observed in CeCoIn$_5$ \cite{CeCoIn5}, YbRh$_2$Si$_2$ \cite{YbRh2Si2}, and YbAgGe \cite{YbAgGe}, while in some other compounds such as CeNi$_2$Ge$_2$ \cite{CeNi2Ge2} and Sr$_3$Ru$_2$O$_7$ \cite{Sr3Ru2O7}, the WF law is verified at the QCP. For the latter one, a sizable residual linear term of thermal conductivity indicates the presence of highly mobile gapless excitations in triangular organics EtMe$_3$Sb[Pd(dmit)$_2$]$_2$ \cite{dmit}, while no magnetic thermal conductivity was observed in another hotly debated QSL candidate YbMgGaO$_4$ \cite{YMGO}.

In this Letter, we report on ultralow-temperature thermal conductivity measurements on single crystals of Pr$_2$Ir$_2$O$_7$. The WF law is verified at high fields and inferred at zero field, suggesting the normal behavior of electrons at the QCP and the absence of fermionic magnetic excitations. A huge magneto-thermal conductivity at finite temperature is found, which may result from the strong scattering of phonons by the fluctuating spins. We shall discuss the implications of these results.

High-quality single crystals of Pr$_2$Ir$_2$O$_7$ were grown by the KF flux methods \cite{KF}. The x-ray diffraction (XRD) measurement determined the largest surface to be the (111) plane (see Supplemental Material \cite{SI}). One Pr$_2$Ir$_2$O$_7$ single crystal (sample A) for electric and thermal conductivity measurements was cut and polished into a rectangular shape of dimensions 0.69 $\times$ 0.38 mm$^2$ in the (111) plane, with a thickness of 0.20 mm. The thermal conductivity was measured in a dilution refrigerator, using a standard four-wire steady-state method with two RuO$_2$ chip thermometers, calibrated $in$ $situ$ against a reference RuO$_2$ thermometer. The electric current and heat current were applied in the (111) plane.

\begin{figure}
\includegraphics[clip,width=6.6cm]{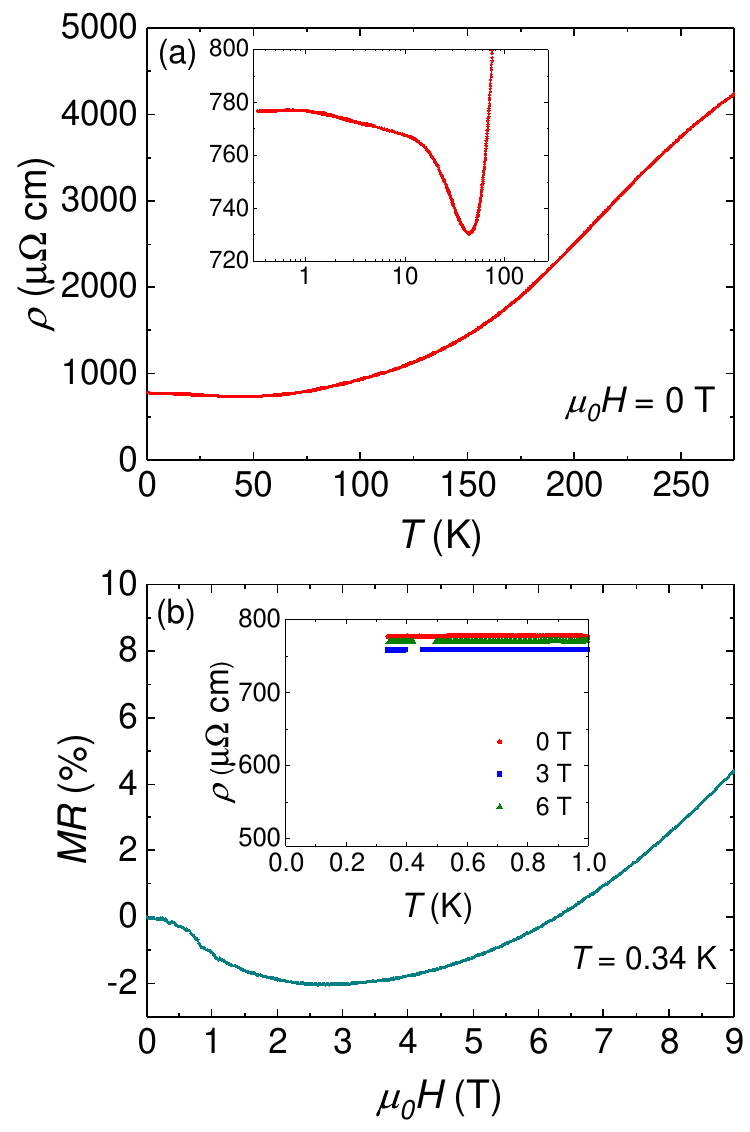}
\caption{(a) Temperature dependence of the resistivity at zero field of Pr$_2$Ir$_2$O$_7$ single crystal. Inset: zoomed view of the resistivity minimum at 45 K due to the Kondo effect. (b) the magnetoresistance at $T$ = 0.34 K. Inset: $\rho(T)$ below 1 K in $\mu_0H$ = 0, 3, and 6 T.}
\end{figure}

\begin{figure}
\includegraphics[clip,width=6.6cm]{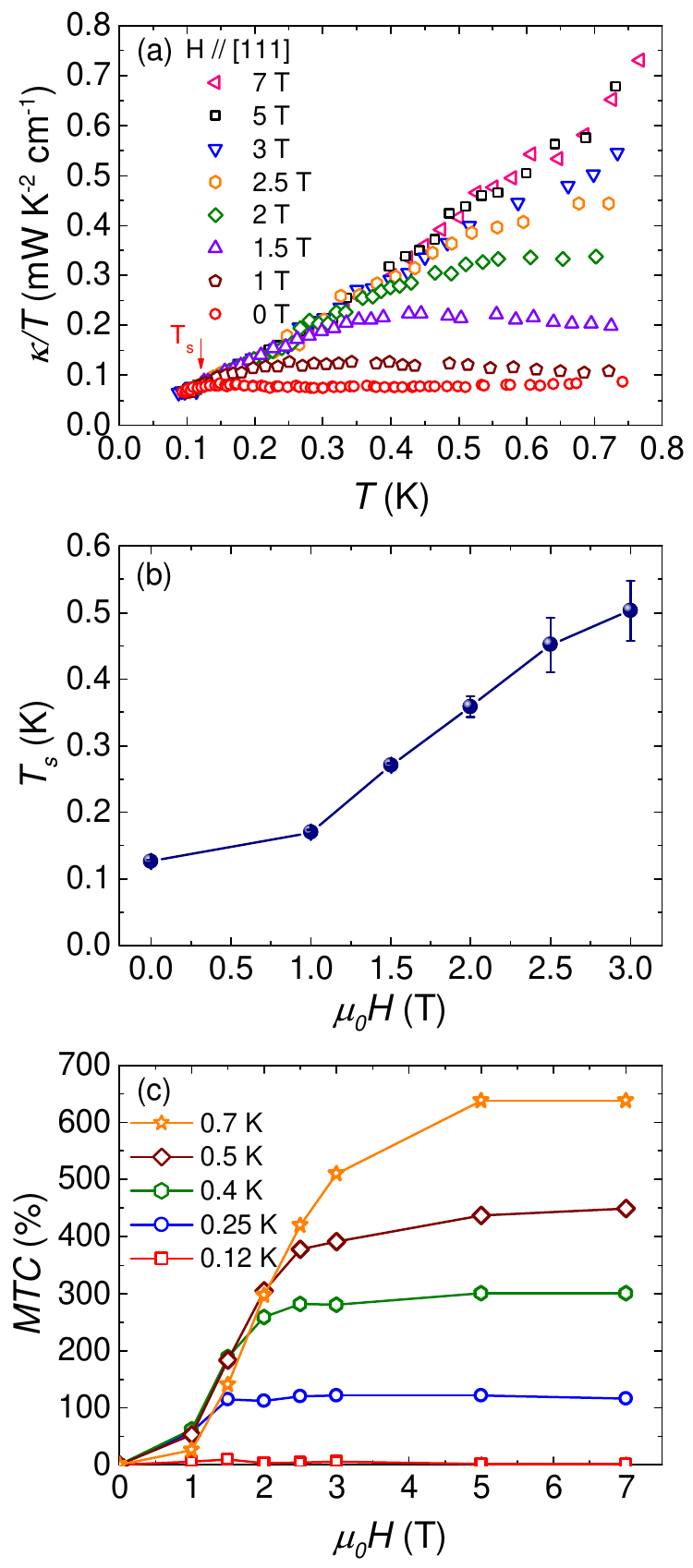}
\caption{(a) The thermal conductivity of Pr$_2$Ir$_2$O$_7$ single crystal at various magnetic fields along the [111] direction. At zero field, the thermal conductivity is strongly suppressed above $T_s \approx$ 0.12 K. (b) Field dependence of the suppression temperature $T_s$. (c) The magneto-thermal conductivity MTC = $(\kappa(H)-\kappa$(0T))/$\kappa$(0T) $\times$ 100 \% at various temperatures. Above 5 T, the thermal conductivity tends to saturate.}
\end{figure}

Figure 1(a) shows the temperature dependence of the resistivity $\rho(T)$ at zero field for the Pr$_2$Ir$_2$O$_7$ single crystal. The Kondo effect is evidenced by the upturn behavior below 45 K where the resistivity displays a minimum, as shown in the inset of Fig. 1(a). This is consistent with Ref. \cite{prl06}. The magnetoresistance MR = ($\rho(H)-\rho$(0T))/$\rho$(0T) $\times$ 100\% at $T$ = 0.34 K is presented in Fig. 1(b). It is quite small, less than 5\% up to 9 T, indicating the little influence of magnetic field on the charge transport. In the inset of Fig. 1(b), $\rho(T)$ below 1 K in $\mu_0H$ = 0, 3, and 6 T are plotted. Since all the curves are very flat, we can safely extrapolate them to the zero-temperature limit and get the residual resistivity $\rho_0$ = 776, 757, and 769 $\mu\Omega$ cm for $\mu_0H$ = 0, 3, and 6 T, respectively.

The thermal conductivities of Pr$_2$Ir$_2$O$_7$ single crystal up to 7 T are shown in Fig. 2(a). The magnetic fields were applied perpendicular to the (111) plane. At high fields like 5 T and 7 T, the thermal conductivity data overlap with each other. With decreasing the field, while $\kappa/T$ data still overlap with the high-field curves below a certain temperature $T_s$, they are suppressed more and more strongly above $T_s$. At zero field, $T_s$ is about 0.12 K. Similar behavior is also observed in another sample B (see the Supplemental Material \cite{SI}). The field-dependence of $T_s$ is plotted in Fig. 2(b).

The magneto-thermal conductivity MTC = $\Delta$$\kappa(H)/\kappa$(0T) = $(\kappa(H)-\kappa$(0T))/$\kappa$(0T) $\times$ 100\% at various temperatures is plotted in Fig. 2(c). MTC tends to saturate above 5 T, when the thermal conductivity curves starts to overlap with each other. In contrast to the magnetoresistance of less than 5\% in charge transport at 0.34 K, the MTC is as large as 100\% at 0.25 K and even 650\% at 0.7 K. For other QSL candidates such as EtMe$_3$Sb[Pd(dmit)$_2$]$_2$ \cite{dmit}, YbMgGaO$_4$ \cite{YMGO}, and $\kappa$-(BEDT-TTF)$_2$Cu$_2$(CN)$_3$ \cite{BEDT}, there is also a positive MTC, but the magnitude is much smaller. We will come back to discuss the origin of this huge MTC later.

\begin{figure}
\includegraphics[clip,width=8cm]{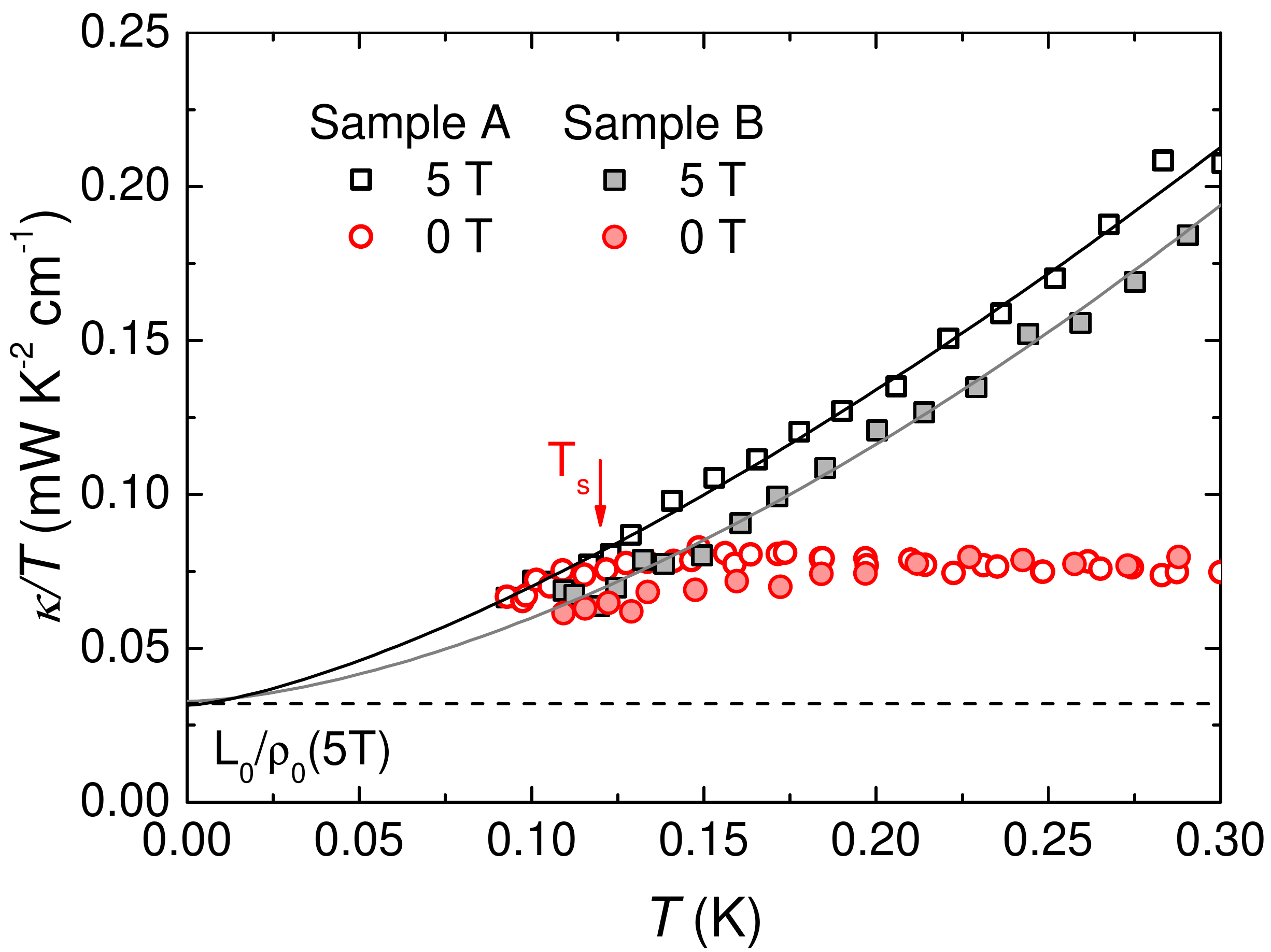}
\caption{The thermal conductivity of two Pr$_2$Ir$_2$O$_7$ single crystals at $\mu_0H$ = 0 and 5 T along the [111] direction, respectively. Solid lines are the fits of the thermal conductivity data to $\kappa/T$ = $a$ + $bT^{\alpha-1}$ at 5 T below 0.3 K. The dashed line is the Wiedemann-Franz law expectation $L_0$/$\rho_0$(5T) = 0.032 mW K$^{-2}$ cm$^{-1}$ for sample A at 5 T, which meets the extrapolated $\kappa_0/T \equiv a$ = 0.031 mW K$^{-2}$ cm$^{-1}$ very well. The overlap of the 0 and 5 T curves below $T_s \approx$ 0.12 K suggests that the Wiedemann-Franz law is also satisfied in zero field.}
\end{figure}

In Fig. 3, we fit the thermal conductivity data below 0.3 K for $\mu_0H$ = 5 T to examine the WF law in Pr$_2$Ir$_2$O$_7$. At ultra-low temperatures, thermal conductivity usually can be fitted to $\kappa/T$ = $a$ + $bT^{\alpha-1}$, where $aT$ represents electrons and other fermionic quasiparticles such as spinons, while $bT^\alpha$ represents phonons and magnons \cite{fit1,fit2}. For phonons, the power $\alpha$ is typically between 2 and 3, due to the specular reflections at the sample surfaces \cite{fit1,fit2}. The fitting gives $\kappa_0/T \equiv a$ = 0.031 $\pm$ 0.008 mW K$^{-2}$ cm$^{-1}$ and $\alpha$ = 2.41 $\pm$ 0.13 for sample A. From Fig. 1(b), $\rho_0$(5T) = 764 $\mu\Omega$ cm is estimated, giving the WF law expectation $L_0/\rho_0$(5T) = 0.032 mW K$^{-2}$ cm$^{-1}$. Therefore, the WF law is satisfied nicely. In order to confirm this result, the data of another sample B are also plotted in Fig. 3. The fitting gives $\kappa_0/T$ = 0.033 $\pm$ 0.006 mW K$^{-2}$ cm$^{-1}$ and $\alpha$ = 2.62 $\pm$ 0.12. Since it has $\rho_0$(5T) = 755 $\mu\Omega$ cm, thus $L_0/\rho_0$(5T) = 0.032 mW K$^{-2}$ cm$^{-1}$, the WF law is also verified in sample B. The verification of WF law above $\mu_0H$ = 5 T is reasonable, because the magnetization approaches saturation \cite{nature10} and the system is away from the QCP \cite{nm14} at high field. As a result, the thermal conductivity above $\mu_0H$ = 5 T is purely contributed from normal electrons and phonons, without other exotic excitations or magnetic scattering.

Since the thermal conductivity data at low fields collapse on the high-field data below $T_s$ (see Fig. 2(a)) and the MR is less than 2\% for $\mu_0H \leq$ 5 T (see Fig. 1(b)), it would be inferred that the WF law is obeyed at all the applied fields, even at zero field. This result is significant. For Pr$_2$Ir$_2$O$_7$, a zero-field QCP was revealed by magnetic Gr\"{u}neisen ratio measurements \cite{nm14}. Violation of the WF law has been observed in CeCoIn$_5$, YbRh$_2$Si$_2$, and YbAgGe at their QCPs \cite{CeCoIn5,YbRh2Si2,YbAgGe}, indicating the breakdown of electrons as Landau quasiparticles. It was interpreted as a consequence of the destruction of Fermi surface \cite{CeCoIn5} or the inelastic scattering \cite{YbRh2Si2,YbAgGe} associated with quantum critical fluctuations. The verification of the WF law in Pr$_2$Ir$_2$O$_7$ at its QCP unambiguously excludes the possibility of the breakdown of Landu quasiparticles and implies the normal formalism of electrons. This is similar to the situation in some other quantum-critical compounds such as CeNi$_2$Ge$_2$ \cite{CeNi2Ge2} and Sr$_3$Ru$_2$O$_7$ \cite{Sr3Ru2O7}.

\begin{figure}
\includegraphics[clip,width=6.6cm]{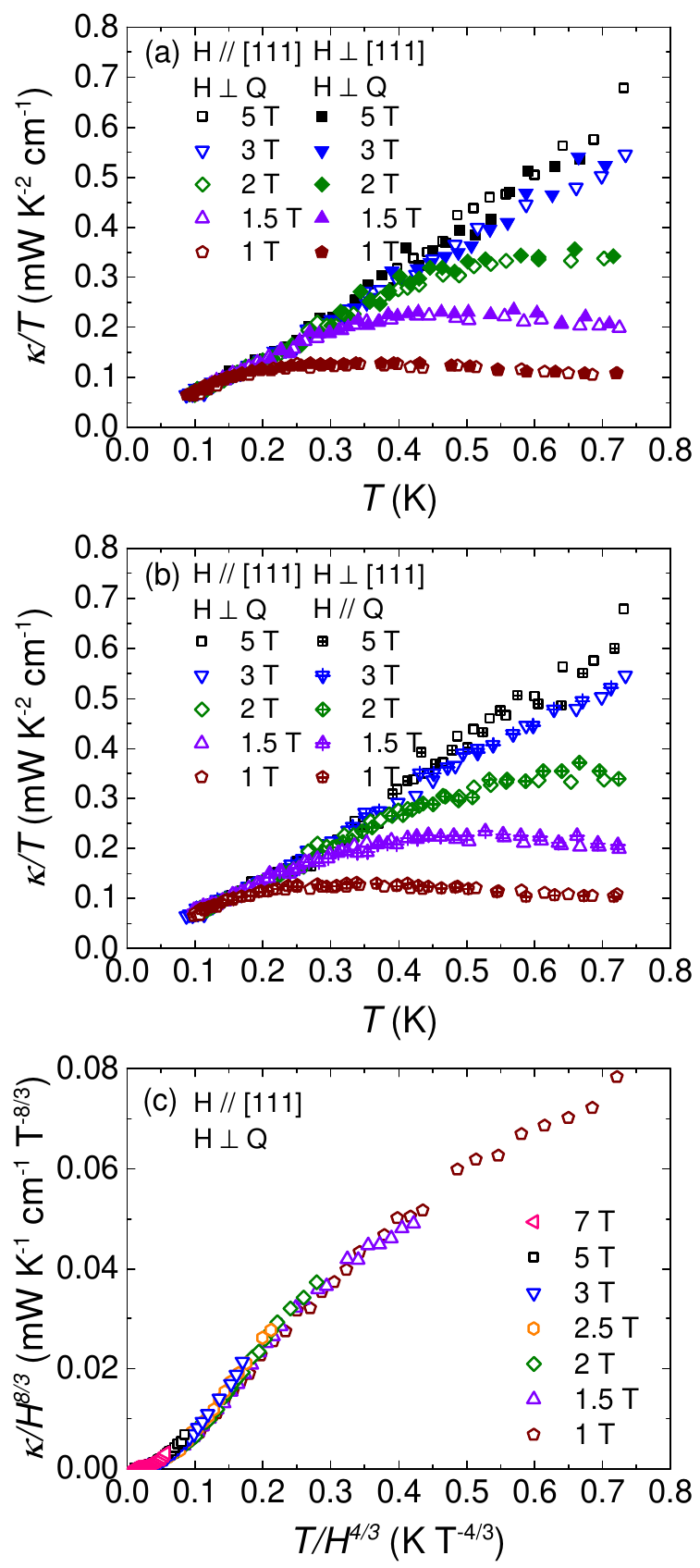}
\caption{(a) The thermal conductivities of Pr$_2$Ir$_2$O$_7$ single crystal in magnetic fields $H$ $\parallel$ [111] and $H$ $\perp$ $Q$ are compared with those in $H$ $\perp$ [111] and $H$ $\perp$ $Q$. (b) They are also compared with those in $H$ $\perp$ [111] and $H$ $\parallel$ $Q$. See Supplemental Material \cite{SI} for schematic illustrations of these three geometries. The overlap of these curves in all three field directions show that the MTC is isotropic. (c) Scaling plot of the thermal conductivity. The plots of $\kappa/H^{8/3}$ versus $T/H^{4/3}$ at various magnetic fields parallel to the [111] direction collapse on the same curve.}
\end{figure}

The verification of WF law in Pr$_2$Ir$_2$O$_7$ demonstrates that there is no additional contribution to the thermal conductivity from mobile fermionic magnetic excitations. Furthermore, since the phonon thermal conductivity in high fields defines the upper boundary of $\kappa$ in Pr$_2$Ir$_2$O$_7$, there is also no positive contribution to $\kappa$ from other bosonic magnetic excitations. For pyrochlore Pr$_2$Ir$_2$O$_7$, one scenario to describe its possible QSL state is the QSI \cite{QSI,condensation}. Three topological excitations, including photon, vison, and magnetic monopole, may emerge from the ground state \cite{QSI,prb04,prx11}. The gapless photons have a rather narrow bandwidth, about 1/1000 of the spin exchange constant $J_{zz}$ \cite{Meng}. Since $J$ of Pr$_2$Ir$_2$O$_7$ is only 1.4 K \cite{nature10}, the photons are likely beyond the accessible temperature regime of our experiment. Both visons and magnetic monopoles have a gap \cite{QSI,prb04,prx11}. It has been claimed that the thermally excited magnetic monopoles contribute to the thermal conductivity in the QSI candidates Yb$_2$Ti$_2$O$_7$ \cite{YTO} and Pr$_2$Zr$_2$O$_7$ \cite{PZO}. However, here in Pr$_2$Ir$_2$O$_7$, we do not observe their positive contribution. If the magnetic monopoles indeed exist in Pr$_2$Ir$_2$O$_7$ with a gap comparable to $J$, the thermally excited magnetic monopoles should be detectable in our temperature range. One possibility is that the velocity and/or mean free path of these excitations may be too small so that their contribution to $\kappa$ is negligible comparing to that of phonons. Another possibility is that the QSI can not describe the ground state of Pr$_2$Ir$_2$O$_7$ thus above-mentioned magnetic excitations do not exist.

Now let us discuss the origin of the huge MTC in Pr$_2$Ir$_2$O$_7$. In Fig. 2(a), starting from the phonon thermal conductivity in high fields, the strong suppression of $\kappa$ at low fields apparently comes from the scattering of phonons by the spin system through the spin-lattice coupling \cite{coupling1,coupling2}, either by well-defined magnetic excitations or by fluctuating spins. In case that there are thermally excited magnetic quasiparticles to scatter the phonons above $T_s$, the field dependence of $T_s$ in Fig. 2(b) suggests that the gap (larger than $T_s$) of these magnetic quasiparticles increases with increasing the field. For above-mentioned magnetic monopoles, the scattering between phonons and monopoles accompanies a spin-flip process, which should be sensitive to the direction of field. Indeed, a metamagnetic transition can be induced only when applying the field along the [111] direction \cite{nature10}. In order to examine this possibility, we measure the thermal conductivity in other two different field directions and compare to the [111] direction, as shown in Figs. 4(a) and 4(b). One can see that the curves in all three field directions overlap with each other. It shows that the MTC is isotropic, i.e., insensitive to the direction of field. Furthermore, the gap of magnetic monopoles should decrease with increasing the field \cite{monopole}, which contrasts to the field dependence of $T_s$. Therefore, the scattering of phonons is unlikely from magnetic monopoles. It will be interesting to theoretically investigate whether there are some other magnetic excitations beyond the QSI scenario. Their gap should be isotropic, and increase with increasing the field.

In case that there are no well-defined magnetic excitations in Pr$_2$Ir$_2$O$_7$, the strong scattering of phonons at low fields may be associated with the fluctuating spins. Intuitively, the spin fluctuations are weakened with lowering the temperature and increasing the field. For Pr$_2$Ir$_2$O$_7$, a bifurcation of the field-cooled and zero-field-cooled magnetic susceptibility curves at about 0.3 K suggests that a partial fraction of spins freezes \cite{prl06,nature10}. Spin freezing has also been observed in classical spin ice like Dy$_2$Ti$_2$O$_7$, and the freezing temperature increases with applying field \cite{DTO,DTO2}. Therefore, this scenario may explain the temperature and field dependence of thermal conductivity in Pr$_2$Ir$_2$O$_7$. In each field, the spins fluctuate very slowly below $T_s$, partially frozen, so that they do not scatter phonons. The field will further weaken the spin fluctuations, thus the $T_s$ increases with increasing field. The spin fluctuations is also isotropic in response to the applied field. Such a simple scenario was used to interpret the thermal conductivity of YbMgGaO$_4$ \cite{YMGO}. One may consider whether it can also apply to the heat transport behavior of other QSI candidates such as Pr$_2$Zr$_2$O$_7$ \cite{PZO}, as recently pointed out by Rau and Gingras in Ref. \cite{RauGingras}.

Interestingly, an unusual scaling behavior $\kappa$ $\sim$ $H^{8/3}$ $F$($T/H^{4/3}$), where $F(x)$ is the scaling function, is observed, as shown in Fig. 4(c). It is also held for sample B (see Supplemental Material \cite{SI}). Note that the scaling law was previously found in the magnetic Gr\"{u}neisen ratio $\Gamma_HH$ versus $T/H^{4/3}$ under 2 T, implying a critical physics in Pr$_2$Ir$_2$O$_7$ \cite{nm14}. Observation of the scaling law in the thermal conductivity is appealing since it is rather rare that the heat transport data scale as a function of a single parameter. Due to the unique metallic nature of Pr$_2$Ir$_2$O$_7$, which is distinct from any other insulating QSL candidates, the interactions between Pr $4f$ moments and Ir $5d$ itinerant electrons may complicate the microscopic description. The scaling law gives a new viewpoint towards such quantum magnets that we hope will stimulate the theoretical study. Other experimental techniques, such as thermal Hall measurement and inelastic neutron scattering, are highly desired to determine the ground state and emergent excitations in this exotic metallic QSL candidate Pr$_2$Ir$_2$O$_7$.

In summary, we have measured the ultralow-temperature thermal conductivity of Pr$_2$Ir$_2$O$_7$ single crystals. The Wiedemann-Franz law is verified at high magnetic fields and inferred at zero field, suggesting the normal behavior of electrons at the zero-field quantum critical point and the absence of mobile fermionic magnetic excitations. A huge isotropic magneto-thermal conductivity is found at finite temperatures, indicating the strong scattering of phonons by the spin system, likely the fluctuating spins. The observed scaling law may help put constraints on the theoretical modelling on Pr$_2$Ir$_2$O$_7$. These results will shed light on the microscopic description on this metallic quantum spin liquid candidate.

We thank Z. Y. Meng, Y. Wan, Y. Zhou for helpful discussions. This work is supported by the Ministry of Science and Technology of China (Grant No: 2015CB921401 and 2016YFA0300503), the Natural Science Foundation of China, the NSAF (Grant No: U1630248), and the fundamental research funds for central universities (2018KFYYXJJ038). \\

\noindent $^\dagger$ E-mail: tianzhaoming$@$hust.edu.cn\\
\noindent $^*$ E-mail: shiyan$\_$li$@$fudan.edu.cn

\clearpage

{\large\bf Supplemental Material for ``Ultralow-temperature thermal conductivity of Pr$_2$Ir$_2$O$_7$: a metallic spin-liquid candidate with quantum criticality''}

\section{I. x-ray diffraction measurement}

\begin{figure}[h!]
\renewcommand\thefigure{S1}
\includegraphics[clip,width=6.6cm]{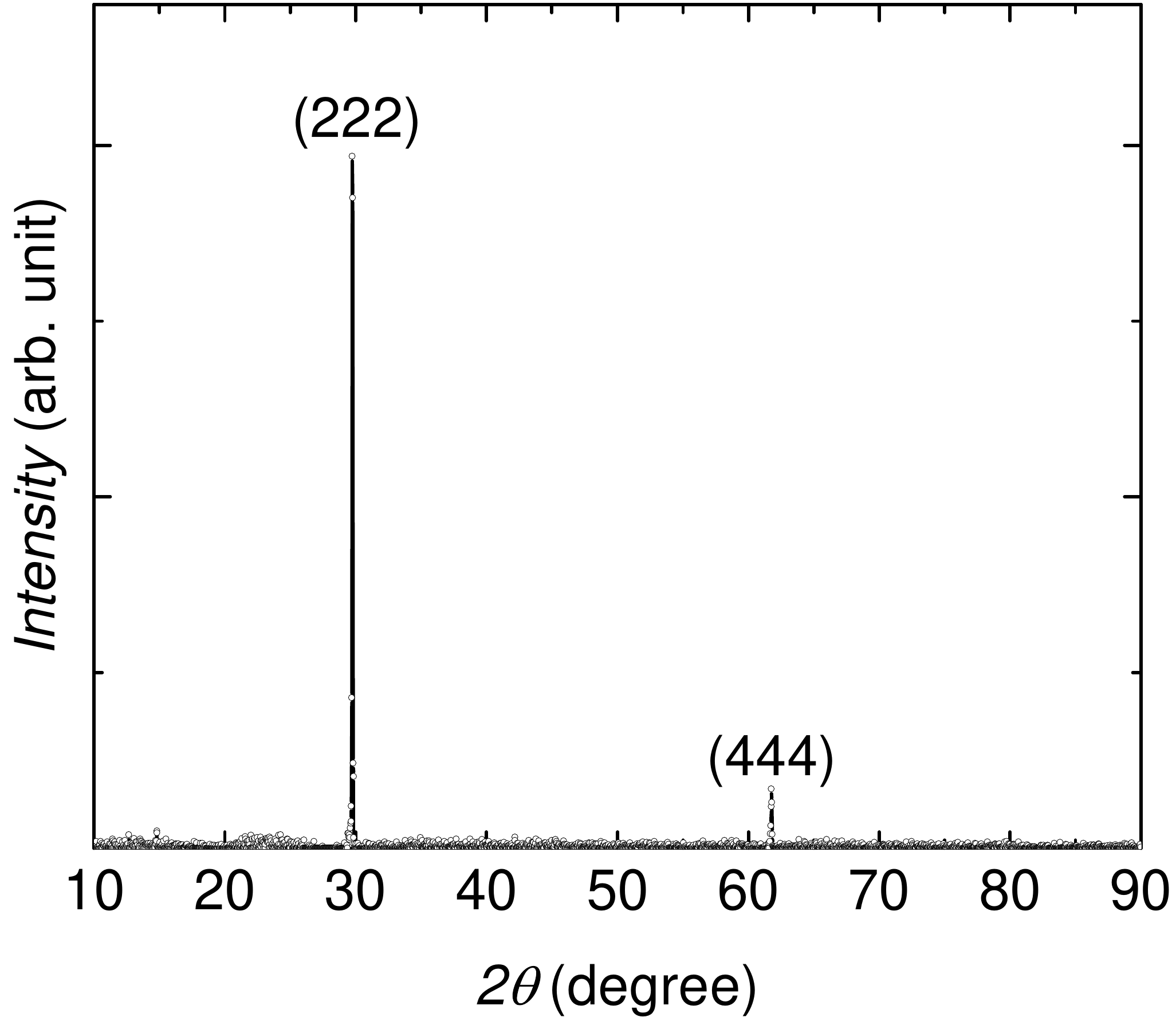}
\caption{Typical x-ray diffraction pattern of the Pr$_2$Ir$_2$O$_7$ single crystal. }
\end{figure}

The typical x-ray diffraction (XRD) data is plotted in Fig. S1, determining the largest surface to be the (111) plane. The heat current is applied in the (111) plane.

\section{II. reproducibility of the transport results}

\begin{figure}[h!]
\renewcommand\thefigure{S2}
\includegraphics[clip,width=6.6cm]{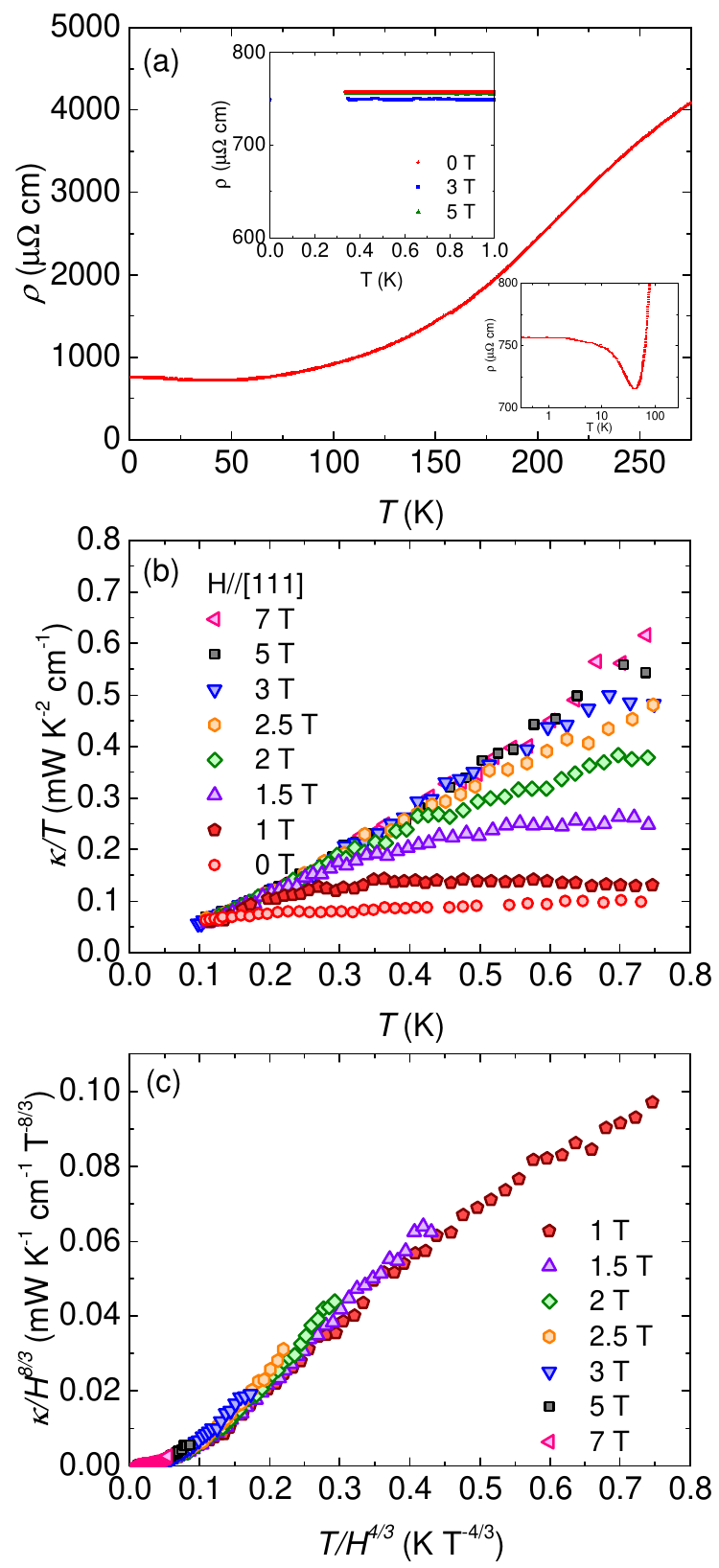}
\caption{(a) Temperature dependence of the resistivity of Pr$_2$Ir$_2$O$_7$ Sample B. Top left inset: $\rho(T)$ below 1 K in $\mu_0H$ = 0, 3, and 5 T. Bottom right inset: zoomed view of the resistivity minimum at 45 K due to the Kondo effect. The magnetic fields were applied along the [111] direction. (b) The thermal conductivity of Sample B at various magnetic fields along the [111] direction. (c) Scaling plot of the thermal conductivity from Sample B. The plots of $\kappa/H^{8/3}$ versus $T/H^{4/3}$ at various magnetic fields parallel to the [111] direction collapse on the same curve.}
\end{figure}

We performed transport measurements on another Pr$_2$Ir$_2$O$_7$ single crystal (Sample B), and obtained similar results to Sample A. Sample B was cut and polished into a rectangular shape with length $l$ = 0.72 mm, width $w$ = 0.26 mm and thickness $t$ = 0.20 mm. The electric and thermal conductivity were measured by standard four-wire method.

Figure S2(a) shows the temperature dependence of the longitudinal resistivity of Sample B at zero field. The Kondo effect is also observed due to the minimum at 45 K (see bottom right inset of Fig. S2(a)), confirming the good quality of our sample. The plots of $\rho(T)$ at 0, 3, and 5 T are presented in the top left inset of Fig. S2(a). By extrapolating to the zero temperature limit, we can get the residual resistivity $\rho_0$ = 757, 749, and 755 $\mu\Omega$ cm for $\mu_0H$ = 0, 3, and 5 T, respectively.

Figure S2(b) shows the thermal conductivities of Sample B up to 7 T. The magnetic fields were applied along the [111] direction. Just as Sample A, the thermal conductivities overlap with each other and deviate from the high-field data above particular temperatures $T_s$. The same scaling law $\kappa$ $\sim$ $H^{8/3}$ $F$($T/H^{4/3}$) is also found in Sample B, see Fig. S2(c).

\begin{figure}[h!]
\renewcommand\thefigure{S3}
\includegraphics[clip,width=8cm]{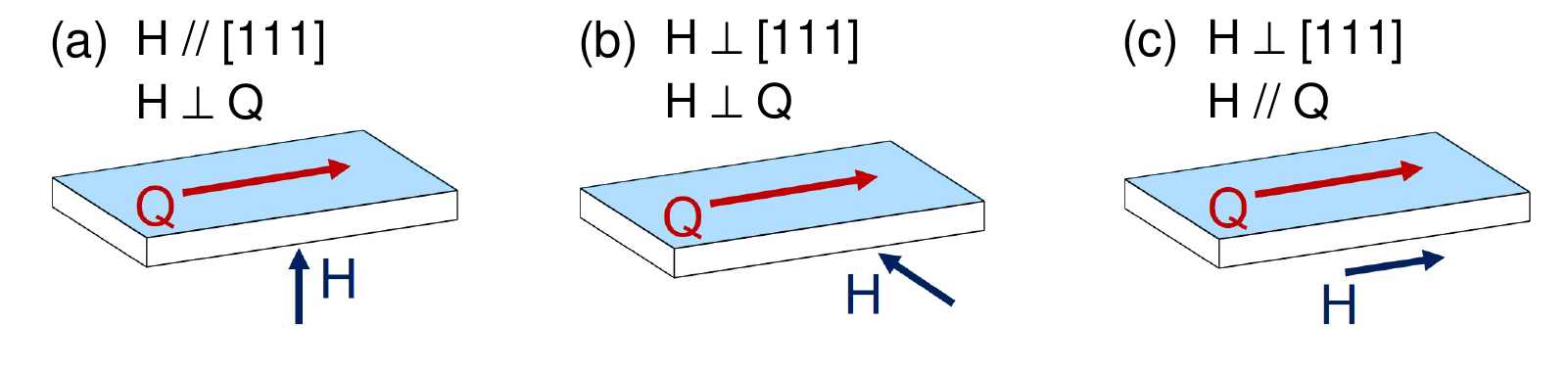}
\caption{(a), (b) and (c) The schematic illustrations of three different relative directions between magnetic field $H$ (blue arrows), heat current $Q$ (red arrows) and [111] direction. The heat currents are all applied in the (111) plane, the largest surface of the sample. The (111) plane is perpendicular to the [111] direction.}
\end{figure}

\section{III. geometries of different magnetic field directions}

In the main part of the paper, the thermal conductivities of Pr$_2$Ir$_2$O$_7$ single crystal in magnetic fields $H$ $\parallel$ [111] and $H$ $\perp$ $Q$ are compared with those in $H$ $\perp$ [111] and $H$ $\perp$ $Q$ (Fig. 4(a) in the main part). They are also compared with the thermal conductivities in $H$ $\perp$ [111] and $H$ $\parallel$ $Q$ (Fig. 4(b) in the main part). These three geometries are illustrated in Fig. S3(a), (b) and (c), respectively.

\end{document}